\documentclass[showpacs,amsmath,twocolumn,showkeys,pra]{revtex4}
\usepackage{dcolumn}    % Align table columns on decimal point
\usepackage{bm}         % bold math
\usepackage{ifpdf}
\usepackage[final]{graphicx}
\usepackage{amssymb,lineno,amsfonts}
\usepackage{graphicx}   % Include figure files
\usepackage{bbm}
\usepackage{mathrsfs}
\usepackage[caption = false,position = top]{subfig}
\usepackage{upgreek}
\usepackage{epstopdf}
\usepackage{setspace}
\usepackage{hyperref}
\usepackage{float}
\usepackage{natbib}
\usepackage{color}
\usepackage{dsfont} 
\usepackage{pdfpages}

\usepackage{tikz}
\usetikzlibrary{shapes.geometric, arrows}
\tikzstyle{startstop} = [rectangle, rounded corners, minimum width=3cm, minimum height=0.8cm,text centered, draw=black, fill=red!30, node distance=1.8cm]
\tikzstyle{io} = [trapezium, trapezium left angle=70, trapezium right angle=110, minimum width=2cm, minimum height=1cm, text centered, draw=black, fill=blue!30]
\tikzstyle{decision} = [diamond, minimum width=3cm, minimum height=1cm, text centered, text width=2.7cm, draw=black, fill=green!30, node distance=1.6cm, aspect=2.5]
\tikzstyle{arrow} = [thick,->,>=stealth]
\tikzstyle{process} = [rectangle, minimum width=2.1cm, minimum height=0.8cm, text centered, text width=2.1cm, draw=black, fill=orange!30, node distance=1.1cm]

\definecolor{med-blue}{RGB}{25,25,112} 
\hypersetup{colorlinks, linkcolor={red},citecolor={blue}, urlcolor={blue}}

\newcommand{\ket}[1]{\vert{#1}\rangle}

\newcommand{\outpr}[2]{\vert{#1}\rangle\langle{#2}\vert}
\newcommand{\inpr}[2]{\langle{#1}\vert{#2}\rangle}
\newcommand{\expv}[1]{\langle{#1}\rangle}

\newcommand{\proj}[1]{\outpr{#1}{#1}}

\newcommand{\tr}{\mathrm{Tr}}

\begin{document}
	
	\title{Robust Quantum Control using Hybrid Pulse Engineering}
	\author{M. Harshanth Ram}
\email{m.harshanthram@students.iiserpune.ac.in}
\affiliation{Department of Physics and NMR Research Center,\\
	Indian Institute of Science Education and Research, Pune 411008, India}
	\author{V. R. Krithika}
\email{krithika\_vr@students.iiserpune.ac.in}
\affiliation{Department of Physics and NMR Research Center,\\
	Indian Institute of Science Education and Research, Pune 411008, India}
	\author{Priya Batra}
\email{priya.batra@students.iiserpune.ac.in}
\affiliation{Department of Physics and NMR Research Center,\\
	Indian Institute of Science Education and Research, Pune 411008, India}
	\author{T. S. Mahesh}
\email{mahesh.ts@iiserpune.ac.in}
\affiliation{Department of Physics and NMR Research Center,\\
	Indian Institute of Science Education and Research, Pune 411008, India}	

	\begin{abstract}
			{
The development of efficient algorithms that generate robust quantum controls is crucial for the realization of quantum technologies. The commonly used gradient-based optimization algorithms are limited by their sensitivity to the initial guess, which affects their performance. Here we propose combining the gradient method with the simulated annealing technique to formulate a hybrid algorithm. Our numerical analysis confirms its superior convergence rate.  Using the hybrid algorithm, we generate spin-selective $\pi$ pulses and employ them for experimental measurement of local noise-spectra in a three-qubit NMR system.  Moreover, here we describe a general method to construct noise-resilient quantum controls by incorporating noisy fields within the optimization routine of the hybrid algorithm.  On experimental comparison with similar sequences obtained from standard algorithms, we find remarkable robustness of the hybrid sequences against dephasing errors.}
\end{abstract}
		


\maketitle

  %%%%%%%%%%%%%%%%%%%%%%%%% Introduction  %%%%%%%%%%%%%%%%%%%%5%%%%%%%%%%%%%%%%
 \section{Introduction}
  \label{Introduction}
 Driven by quantum technology goals, the control of quantum mechanical systems has become an important topic of research in the recent
years. 
Robust quantum control lies in the cornerstone of any efficient and reliable quantum processor.  The building blocks of quantum memory, the qubits, must be precisely controlled and protected against systematic deviations in the control fields, as well as against the random noise induced by the surrounding environment. 
Accordingly, a plethora of control techniques have been developed, which include  gradient-based approaches such as strongly modulated pulses
\cite{fortunato2002design, mahesh2006quantum}, gradient ascent pulse
engineering (GRAPE) \cite{14, 15}, gradient optimization of analytical control \cite{16},  truncated basis approaches such as
chopped random basis optimization \cite{20, 21}, variational-principle-based techniques like relaxation optimised pulse engineering \cite{khaneja2003optimal},  Krotov
optimization \cite{17, 18, 19}, a combination of gradient and variational controls like K-BFGS algorithm \cite{eitan2011optimal}, evolutionary algorithm based controls \cite{zahedinejad2014evolutionary, bhole2016steering}, as well as neural network  and reinforcement learning inspired approaches \cite{niu2019universal, an2019deep}. 

Quantum control techniques have been realised in Nuclear Magnetic Resonance (NMR) \cite{ryan2008liquid, negrevergne2006benchmarking, vandersypen2005nmr, sun2014experimental}, nitrogen-vacancy centers \cite{golter2016optomechanical},  superconducting qubits \cite{dong2015robust}, ion traps \cite{PhysRevLett.91.157901}, magnetic resonance imaging \cite{grinolds2011quantum}, and cold atoms \cite{chu2002cold}.
Here we use NMR systems as quantum platforms to develop robust quantum control techniques. The availability of long-lasting spin coherences and highly adaptable control fields make NMR an ideal testbed for quantum control developments. 

      \begin{figure}
% 	 \centering
   		\includegraphics[trim=3cm 4cm 5cm 3.5cm,width=7cm,clip=]{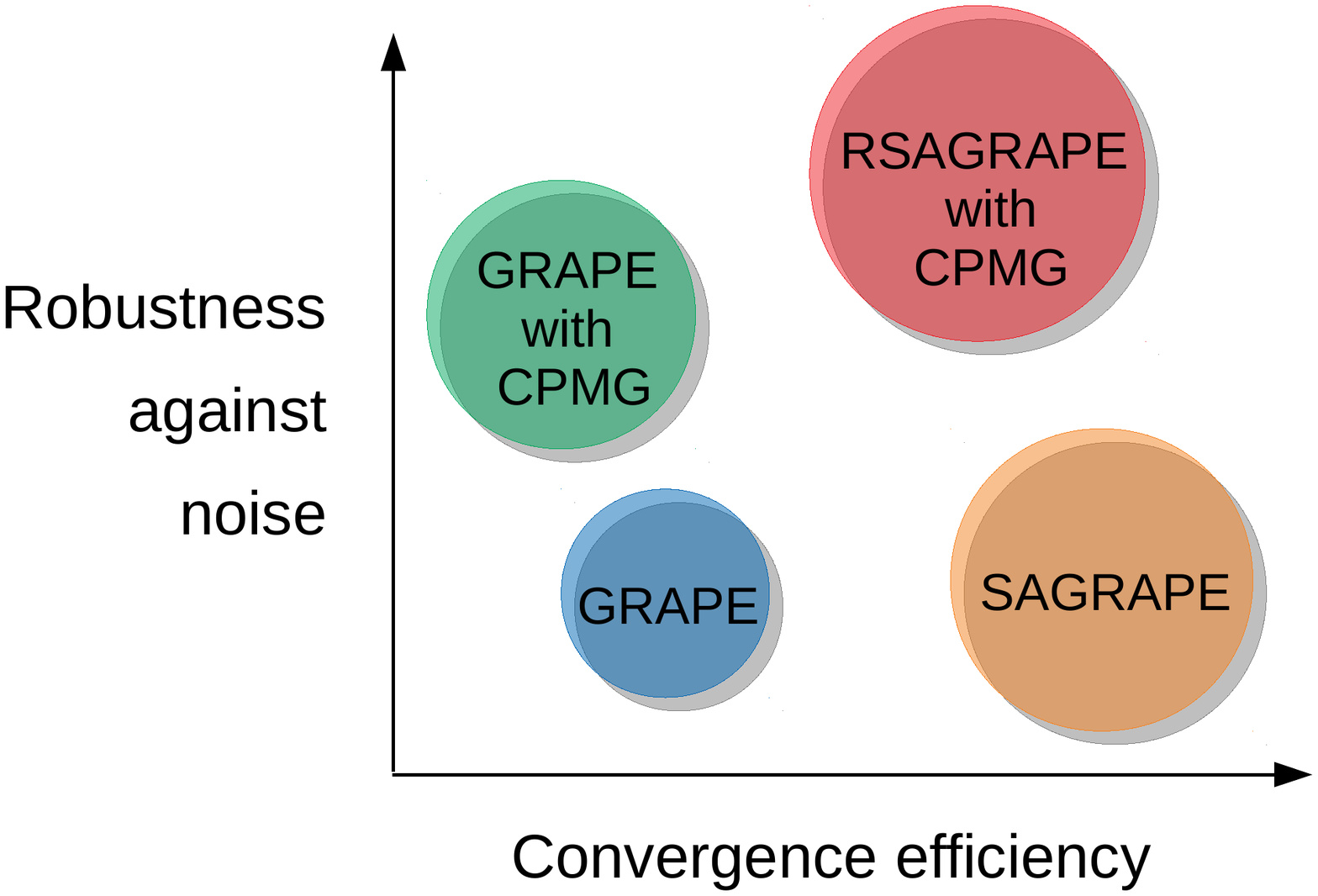}
   	\caption{Comparison of quantum control algorithms w.r.t. convergence efficiency and robustness against noise.}
   	\label{overview}
   \end{figure}

Generally quantum control techniques are limited by two challenges:
(i) sensitivity to initial guess as well as local optima which limit the convergence efficiency of the optimization algorithms, and (ii) susceptibility to lose fidelity due to incoherence, decoherence, or a combination of both.  In this article, we propose methodologies to combat both of these limitations (see Fig. \ref{overview}).  First, we propose combining GRAPE and simulated annealing (SA) \cite{Simul,Simulatedannealinggrape} to realize a hybrid optimization technique called SAGRAPE that can overcome local optima and converge faster toward better solutions. 
Secondly, we propose optimizing SAGRAPE sequences along with certain random fields, yielding a robust algorithm, namely RSAGRAPE. It can generate control sequences that are resilient against environmental noises.  Additionally, we describe integrating these sequences with standard dynamical decoupling sequences such as CPMG \cite{carr1954effects,meiboom1958modified}, which enhances their robustness.  We numerically analyze the convergence efficiency and experimentally demonstrate the robustness of the hybrid sequences.

This article is organized as follows. In Sec. \ref{sec:Simuanneal}, we describe the quantum system and revisit the GRAPE  algorithm.  Subsequently, we introduce the hybrid algorithm SAGRAPE and analyze its convergence efficiency.  We then describe the NMR implementation of SAGRAPE sequences. In Sec.~\ref{sec:Hamiltonian} we describe SAGRAPE optimization in the presence of a random field to obtain RSAGRAPE, and experimentally demonstrate the robustness of RSAGRAPE sequence.  Finally we conclude in  Sec.~\ref{sec:conclusions}.

%%%%%%%%%%%%%%%%%%%%%%%%%%%%%%%%%%%%%%%%%%%%%%%%%%%%%%%%%%%%%%%%%%%%%%%%%%%%%%%

 \section{GRAPE and SAGRAPE}
    \label{sec:Simuanneal}
In this section, we first describe the quantum system on which we are interested in establishing quantum control, both in terms of state to state transfer, as well as realizing general quantum gates.  After reviewing the GRAPE as well as the SA algorithms, we explain the hybrid algorithm SAGRAPE and analyze its convergence efficiency.  Finally, we describe an NMR application of SAGRAPE sequences.

\subsection{The quantum system}
We consider an $n$-qubit NMR system with a total rotating-frame Hamiltonian under the weak-coupling approximation,
\begin{align}
H(t) &= H_{0} + H_\mathrm{RF}(t),
~~ \mbox{where} \nonumber 
\\ H_{0}&=-\sum_k\Omega_k I_{kz} + 2\pi \sum_{kl} J_{kl} I_{kz}I_{lz},
\label{eq:HandH0}
\end{align}
where $\Omega_k$ are the resonance offsets, $J_{kl}$ is the indirect spin-spin coupling constant, and $I_{k\alpha}$ with $\alpha \in \{x,y,z\}$ are components of the $k$th spin operator $\mathbf{I}_k$.  
Generally, we discretize the entire sequence of duration $T$ into $N$ equal segments each of duration $\tau = T/N$.
We assume that the $n$ spin-qubits belong to $s$ different species (isotopes) and the $q$th species is controllable with a RF sequence $\{\omega_{q\alpha}(j)\}$. 
The RF Hamiltonian for $j$th segment is of the form
\begin{align}
H_\mathrm{RF}(j) 
&= \sum_{q=1}^s \sum_{\alpha = x,y} \omega_{q\alpha}(j) H_{q\alpha}
\nonumber \\
~~\mbox{with}~~
H_{q\alpha} &= \sum_{k \in \{\mathrm{species}~q\}} I_{k\alpha}.
\end{align}
In this work, we consider only one type of nuclear species, i.e., $q = 1$, and hence we drop the subscript $q$ from now onward.

In practice, there exists a spatial inhomogeneity of RF amplitudes which is modeled by a scaling factor $r_{m}$ with an associated probability $p_m$.  Thus for the $m$th subensemble, the total Hamiltonian of $j$th segment is
\begin{align}
H^m(j) &= H_0 + H^m_\mathrm{RF}(j), 
~~\mbox{with}
\nonumber \\
H^m_\mathrm{RF}(j) 
&=  r_m\left\{\omega_{x}(j) H_x +\omega_{y}(j) H_y\right\}.
\label{eq:HandHrf}
\end{align}
The propagator for the entire sequence is written as
\begin{align}
U^m = \prod_{j=1}^{N} U_j^m, ~~\mbox{where}
~~
U_j^m = \exp\left(
-i\tau H^m(j)
\right)
\end{align}
are the segment propagators.
The aim of the control algorithms is to generate the sequence $\{\omega_x(j),\omega_{y}(j)\}$ that perform a given quantum control task.  In the following we revisit the GRAPE algorithm that is commonly used for this purpose.

\subsection{Gradient Ascent Pulse Engineering (GRAPE)}
Given a target operation, GRAPE starts with a random sequence $\{\omega^0_x(j),\omega^0_{y}(j)\}$, which is updated based on the local gradients.  Generally, we consider the following two types of quantum control.

\subsubsection{State control}
The objective here is find a sequence that transforms an initial state $\rho_0$ to a target state $\rho_F$ by finding a sequence that maximizes the average state fidelity
\begin{align}
\Phi = \sum_m p_m \inpr{\rho_F}{U^m\rho_0U^{m\dagger}}
=\sum_m p_m \tr\left[\rho_F U^m\rho_0U^{m\dagger}\right],
\label{statefid}
\end{align} 
where the summation is carried out on the subensembles corresponding to RF inhomogeneity (RFI).
The update rule for the $i$th iteration is \cite{14}
\begin{align}
      \omega_\alpha^{i+1}(j) &= \omega_\alpha^{i}(j) + \epsilon   G_\alpha^{i}(j), ~~\mbox{where the gradient} 
      \nonumber \\
      G_\alpha^{i}(j) &= 
       -\iota \tau \sum_m p_m \left\langle\lambda^{mi}_{j} \vert \left[H_\alpha, \rho_{j}^{mi}\right]\right\rangle, ~~\mbox{where}  \nonumber \\
      \rho^{mi}_{j} &=U^{mi}_{j} \cdots U^{mi}_{1}\ \rho_{0}  U^{mi\dagger}_{1} \cdots U^{mi\dagger }_{j}, ~~\mbox{and}
       \nonumber \\
  \lambda^{mi}_{j} &=U^{mi\dagger  }_{j+1} \cdots U^{mi\dagger}_{N}\ \rho_{F} U^{mi}_{N} \cdots U^{mi}_{j+1}.
  \end{align}
  Here $\epsilon$ is the step-size.  The iterations are continued until the  fidelity $\Phi$ reaches the desired value.

\subsubsection{Gate control:}
Here the objective is to generate a sequence that realizes a desired propagator $U_F$ by maximizing the gate fidelity
\begin{align}
\Phi =\sum_{m}p_{m} |\inpr{U}{U_F}|^2
=\sum_{m}p_{m} \left\vert\tr\left[U^\dagger U_F\right]\right\vert^2.
\label{gatefid}
\end{align} 
The update rule for the $i$th iteration is \cite{14}
  \begin{align}
\omega_\alpha^{i+1}(j) &= \omega_\alpha^{i}(j) + \epsilon G_\alpha^{i}(j), ~~\mbox{where the gradient} 
      \nonumber \\
      G_\alpha^{i}(j) &= 
      -2\iota \tau \sum_{m}p_{m}\mathrm{Re}\left\{\left\langle P^{mi}_{j} \mid H_\alpha X^{mi}_{j}\right\rangle\left\langle X^{mi}_{j} \mid P^{mi}_{j}\right\rangle\right\},       \nonumber \\
\mbox{with}  ~
X^{mi}_{j}&=U^{mi}_{j}\ U^{mi}_{j-1} \cdots U^{mi}_{2}\ U^{mi}_{1},~\mbox{and}
\nonumber
\\
P^{mi}_{j}&=U^{mi\dagger }_{j+1}\ U^{mi\dagger }_{j+2}\cdots U^{mi\dagger }_{N-1}\ U^{mi\dagger }_{N}\ U_{F}.   \end{align}
  where $\epsilon$ is the step-size.  Again, the iterations are continued until the  fidelity $\Phi$ reaches the desired value.
  
        \begin{figure}
% 	 \centering
   		\includegraphics[trim=0cm 0cm 0cm 0cm,width=8cm,clip=]{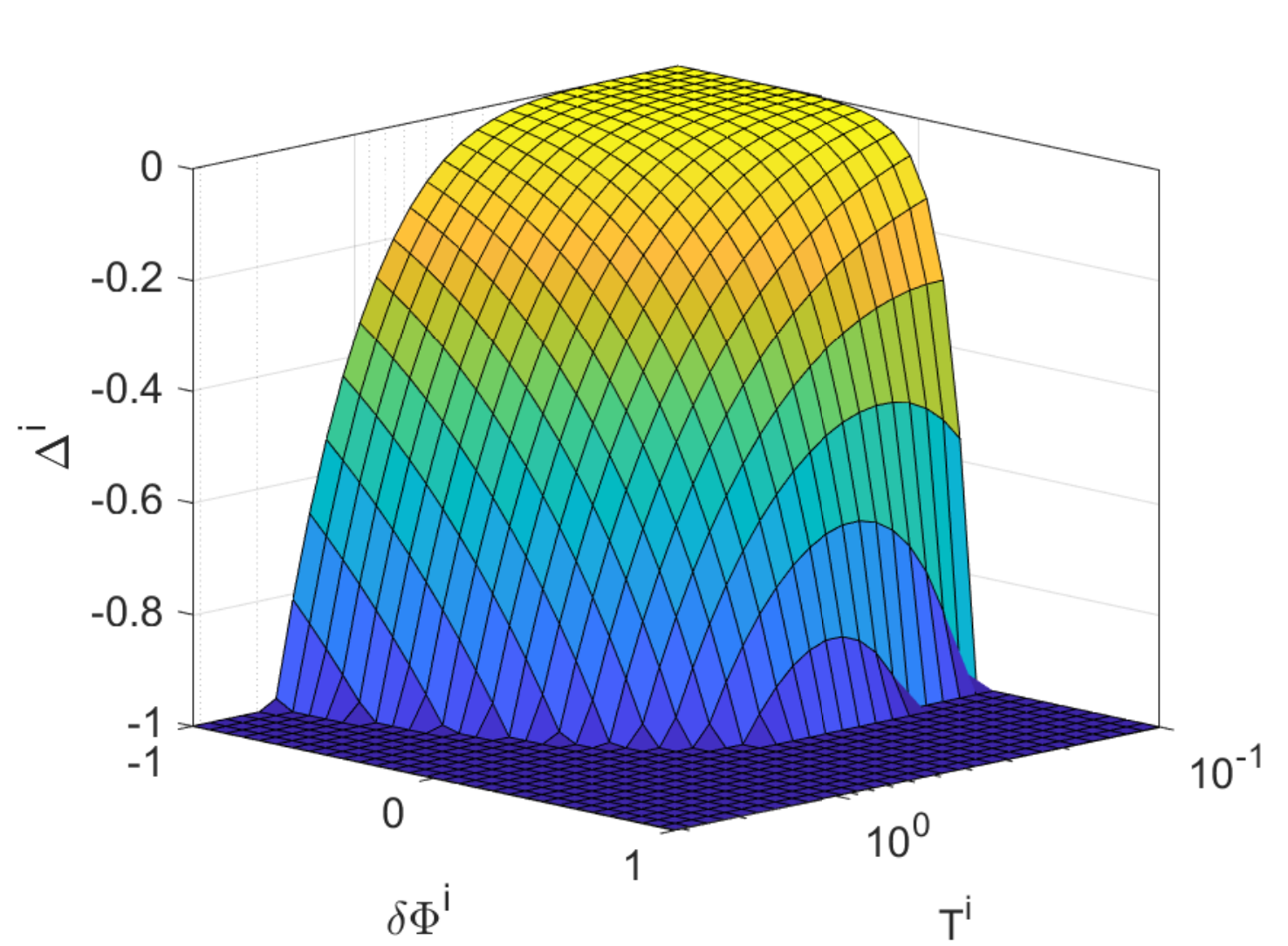}
   	\caption{ Threshold function ($\Delta^{i}$) plotted versus  temperature T$^{i}$ and the fidelity variation $\delta\Phi^{i}$. Note that $\Delta^{i}$ is bounded by $[-1,0]$.}
   	\label{figureone}
   \end{figure}
    
    \subsection{Simulated annealing (SA)}
    SA is a single candidate-based probabilistic algorithm that is used to reach the neighbourhood of the global maximum of the fitness function. 
    The general idea of SA can be described as two modes - an exploration mode and an exploitation mode as explained in the following.    In every iteration, a solution $\{\omega_\alpha'\}$ of fidelity $\Phi(\{\omega_\alpha'\})$ is randomly selected from the search-space from the neighborhood of the current solution $\{\omega_\alpha^i\}$ with fidelity $\Phi(\{\omega_\alpha^i\})$.  The current solution is replaced with the random solution if the fidelity variation,
    \begin{align}
       \delta\Phi^i &= \Phi(\{\omega_\alpha'\})-\Phi(\{\omega_\alpha^i\})  \ge \Delta^i, 
       \nonumber \\
       \mbox{where}~ \Delta^i &= -\min\left[1,\mbox{T}^{i}\exp \left(\frac{\delta\Phi^i}{\mbox{T}^{i}}\right)\right]. 
       \label{fidvar}
    \end{align}
    Here T$^{i}$ is referred to as the temperature of the current iteration in analogy to the thermodynamical processes.
    If $\delta\Phi^{i} \ge 0$, then the fidelity has improved with the random solution, and naturally
    $\{\omega_\alpha^{i+1}\}$ is set to $\{\omega_\alpha'\}$.  However, notice that even if $\delta\Phi^{i} < 0$, i.e., the random solution is worse than the current solution, then we may still set
    $\{\omega_\alpha^{i+1}\}$ to $\{\omega_\alpha'\}$ with a non-zero probability. Thus the algorithm is in a exploration mode, where it looks for a neighbourhood with favorable solutions. This is the salient feature of the simulated annealing algorithm that enables it to get over the local optima. As explained above, the temperature parameter controls the threshold function  for selecting nonoptimal solutions for subsequent iterations. Higher the temperature, greater are the chances of the nonoptimal solution to become the next solution, but as the algorithm approaches the global optimum, it should promote only candidates that increase the fidelity. Thus, in the initial iterations of SA, the temperature is kept high and as iterations pass, the  temperature is gradually reduced. This is achieved by multiplying the temperature with a cooling factor ($\gamma < 1$ ) at every iteration.  Accordingly, the algorithm gradually shifts from exploration mode to exploitation mode.  This is illustrated in Fig. \ref{figureone}, wherein the threshold function is plotted versus $\delta\Phi^{i}$ as well as temperature $\mbox{T}^{i}$.

        \begin{figure}
{\footnotesize
\begin{tikzpicture}[node distance=2cm]
\node (start) [startstop] {Randomly choose $\{\omega_\alpha^0\}$};
\node (pro0) [process, below of=start, xshift = 4cm] {Set $T^{0}$ \& $\gamma$};
%\node (pro0) [process, below of=start] {Set $T^{0}$ \& $\gamma$};
\node (pro1) [process, below of=pro0, xshift = -4cm] {Randomly choose $\{\omega_\alpha'\}$};
\node (pro1a) [process, below of=pro1] {Find $\delta\Phi^i$; set $T^{i+1} = \gamma T^i$};%\node (pro1) [process, below of=in1] {Process 1};
\node (dec1) [decision, below of=pro1a] {Is  $\delta\Phi^i \ge \Delta^i$?};
\node (pro2a) [process, below of=dec1, yshift=-0.5cm] {Set $\{\omega_\alpha^{i+1}\} = \{\omega_\alpha'\}$};
\node (pro2b) [process, right of=pro2a, xshift=1.5cm] {Set $\{\omega_\alpha^{i+1}\}=\{\omega_\alpha^{i}\}$};
\node (dec2) [decision, below of=pro2a] {Is  $i < \kappa$?};
\node (pro3a) [process, below of=dec2, yshift=-0.5cm] {Apply GRAPE on $\{\omega_\alpha^\kappa\}$};
\node (pro3b) [process, left of=pro2a, xshift=-1.5cm] {Set $T^{i+1} = \gamma T^i$};
\node (dec2) [decision, below of=pro2a] {Is  $i < \kappa$?};
\node (pro3a) [process, below of=dec2, yshift=-0.5cm] {Apply GRAPE on $\{\omega_\alpha^\kappa\}$};
\node (dec3) [decision, below of=pro3a] {Finished SAGRAPE iterations?};
\node (stop) [startstop, below of=dec3] {Stop};

\draw [arrow] (start) -| (pro0);
\draw [arrow] (pro0) -| (pro1);
\draw [arrow] (pro1) -- (pro1a);
\draw [arrow] (pro1a) -- (dec1);
\draw [arrow] (dec1) -- node[anchor=east] {yes} (pro2a);
\draw [arrow] (dec1) -| node[anchor=south] {no} (pro2b);
\draw [arrow] (dec2) -- node[anchor=east] {no} (pro3a);
\draw [arrow] (dec2) -| node[anchor=east] {yes} (pro3b);

\draw [arrow] (pro3b) |- (pro1);
\draw [arrow] (pro2a) -- (dec2);
\draw [arrow] (pro2b) |- (dec2);

\draw [arrow] (pro3a) -- (dec3);
\draw [arrow] (dec3) -| node[anchor=north] {no} (pro0);
\draw [arrow] (dec3) -- node[anchor=east] {yes} (stop);
% \draw [arrow] (dec3) -- node[anchor=east] {yes} (pro3a);
\end{tikzpicture}
}        
        
% 	 \centering
% 	\includegraphics[trim=0cm 0cm 0cm 0cm,width=8cm,clip=]{exp_report/manuscript/hybridalgorithmchart.JPG}
  	\caption{Flowchart for the  SAGRAPE algorithm.}
  	\label{figuretwo}
  \end{figure}
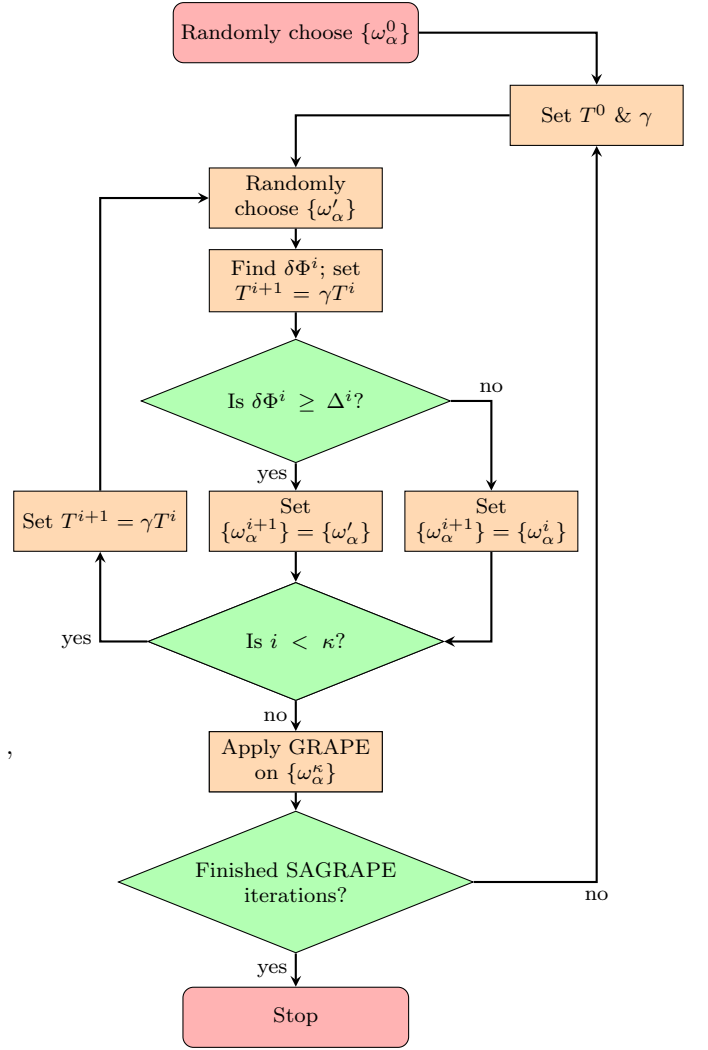

\subsection{SAGRAPE algorithm}
Even though SA is efficient in getting out of the local optima and identifying a good neighbourhood, it takes many iterations to get to the global optimum. On the other hand, a gradient based algorithm like GRAPE \cite{14} is much faster in identifying the best solution once a good neighbourhood is reached. Thus, we now introduce $\kappa$ iterations of SA before each GRAPE iteration to create one iteration of SAGRAPE algorithm. The solution from the GRAPE algorithm of the current SAGRAPE iteration is chosen as the initial solution for the SA algorithm of the next SAGRAPE iteration.   
This way, we can incorporate the best of both the optimization techniques.  The flowchart for the SAGRAPE algorithm is shown in Fig. \ref{figuretwo}.

 \begin{figure}%
    \subfloat[]{
    {\includegraphics[trim=0cm 0cm 0cm 0.7cm,width=8cm,clip=]{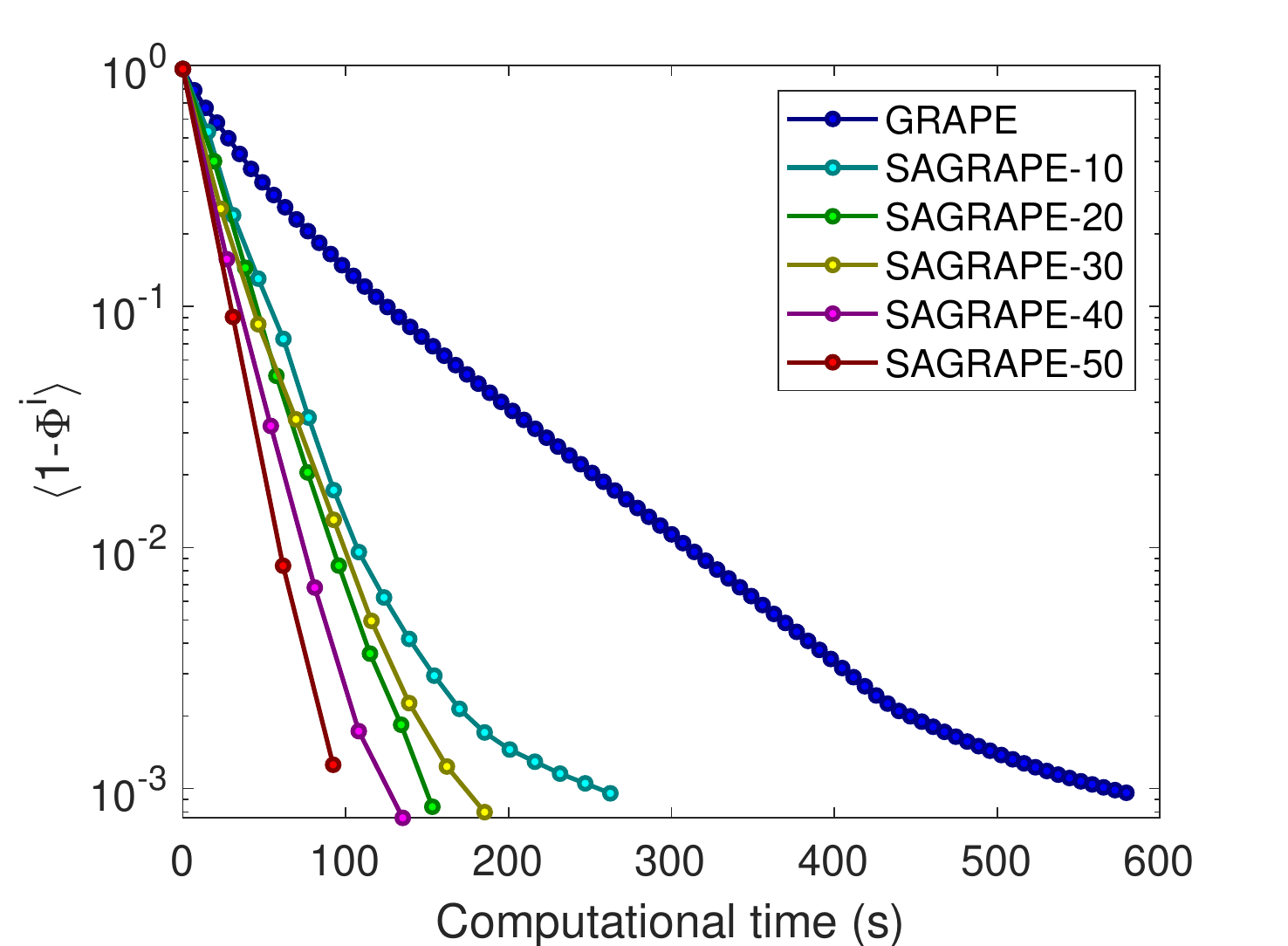}}}%
    \qquad
    \subfloat[]{
   {\includegraphics[trim=0cm 0cm 0cm 0.7cm,width=8cm,clip=]{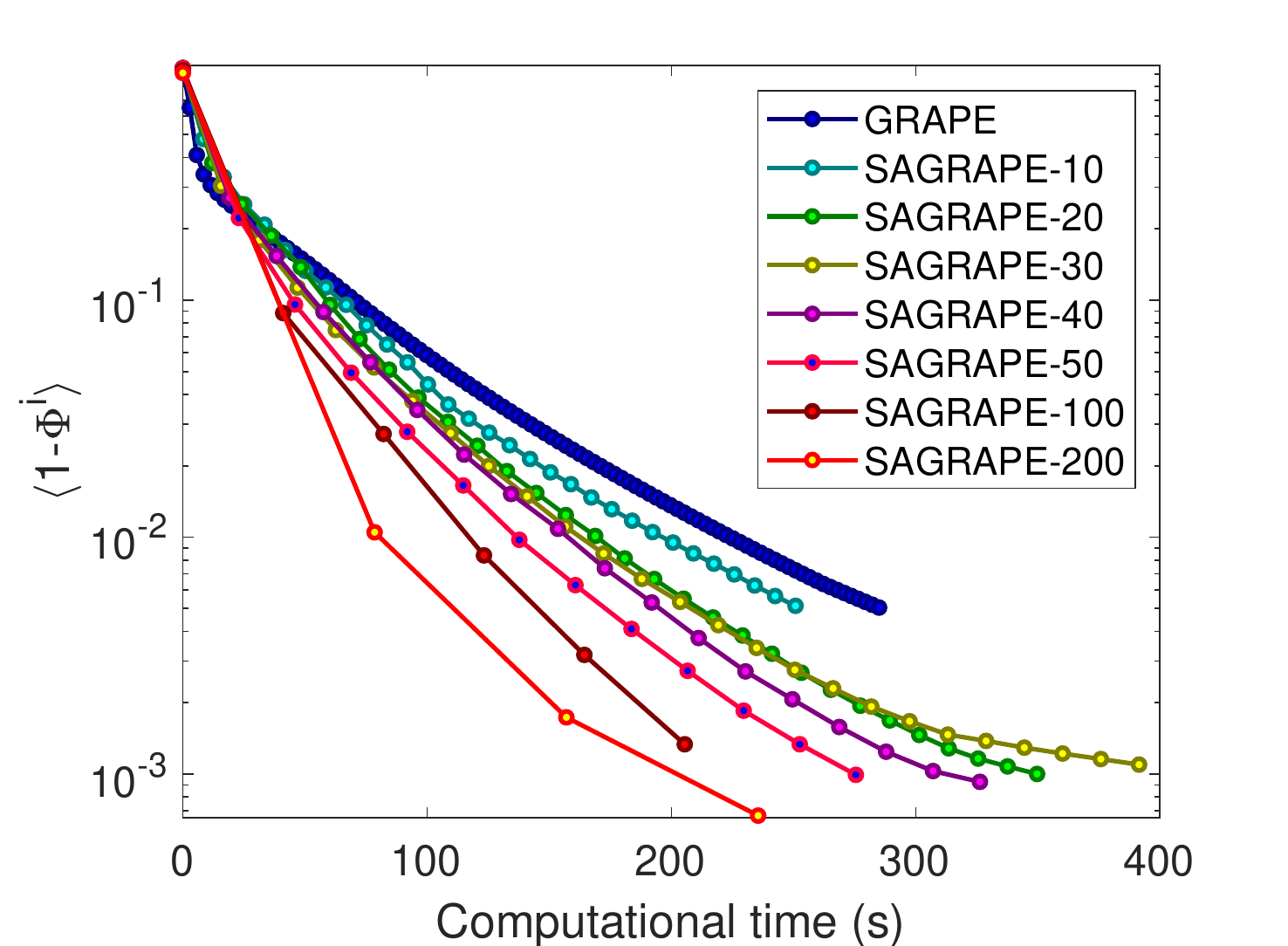} }}%
    \caption{Average infidelity $\expv{1-\Phi^i}{}$ versus computational time taken by GRAPE and SAGRAPE-$\kappa$ algorithms for (a) state control  and (b) gate control  respectively.  
    %Here $\kappa$ indicates the number of SA iterations carried out before every GRAPE iteration.  
    Each SAGRAPE-$\kappa$ data point corresponds to $\kappa$ number of SA iterations followed by one GRAPE iteration.
    }%
    \label{hybriditerationsresult}%
\end{figure}

\subsection{Convergence analysis of SAGRAPE}
We now demonstrate the convergence of the SAGRAPE algorithm for state control and gate control in a two qubit system. For state control, we generate the RF amplitudes $\{\omega_{\alpha}(j)\}$ that transform the two-spin NMR thermal equilibrium state to the long-lived singlet state (LLS), i.e.,
\begin{align}
I_{z1} + I_{z2}
\stackrel{\{\omega_{\alpha}(j)\}}{\longrightarrow}-\mathbf{I}_{1}\cdot \mathbf{I}_{2}.
\end{align}
Similarly, for the gate control, we generate the sequence realizing the CNOT gate
\begin{align}
U = \proj{0} \otimes \mathbbm{1}
+ \proj{1} \otimes \sigma_x,
\end{align}
where $\sigma_x$ is the Pauli-x operator which applies the NOT gate on the second qubit only if the first qubit is in state $\ket{1}$.

All the  sequences prepared with GRAPE and SAGRAPE algorithm have the same time duration of 120 ms discretized into 600 equal-duration segments.  We carried out the convergence analysis of `SAGRAPE-$\kappa$' with varying numbers of SA iterations $\kappa$ and  compared the computational time with the GRAPE algorithm.
The results are shown in Fig. \ref{hybriditerationsresult}.
Here each curve is obtained by averaging over five different trials starting from random initial guess $\{\omega_\alpha^0(j)\}$.
It is evident that SA significantly improves the convergence efficiency in both state and gate control tasks.

%%%%%%%%%%%%%%%%%%%%%%%%%%%%%%%%%%%%%%%%%%%%%%%%%%%%%%%%%

\subsection{NMR demonstration of SAGRAPE}
\begin{figure}%
    \subfloat[]{
    \includegraphics[trim=3cm 5cm 3cm 4.4cm,width=5.5cm,clip=]{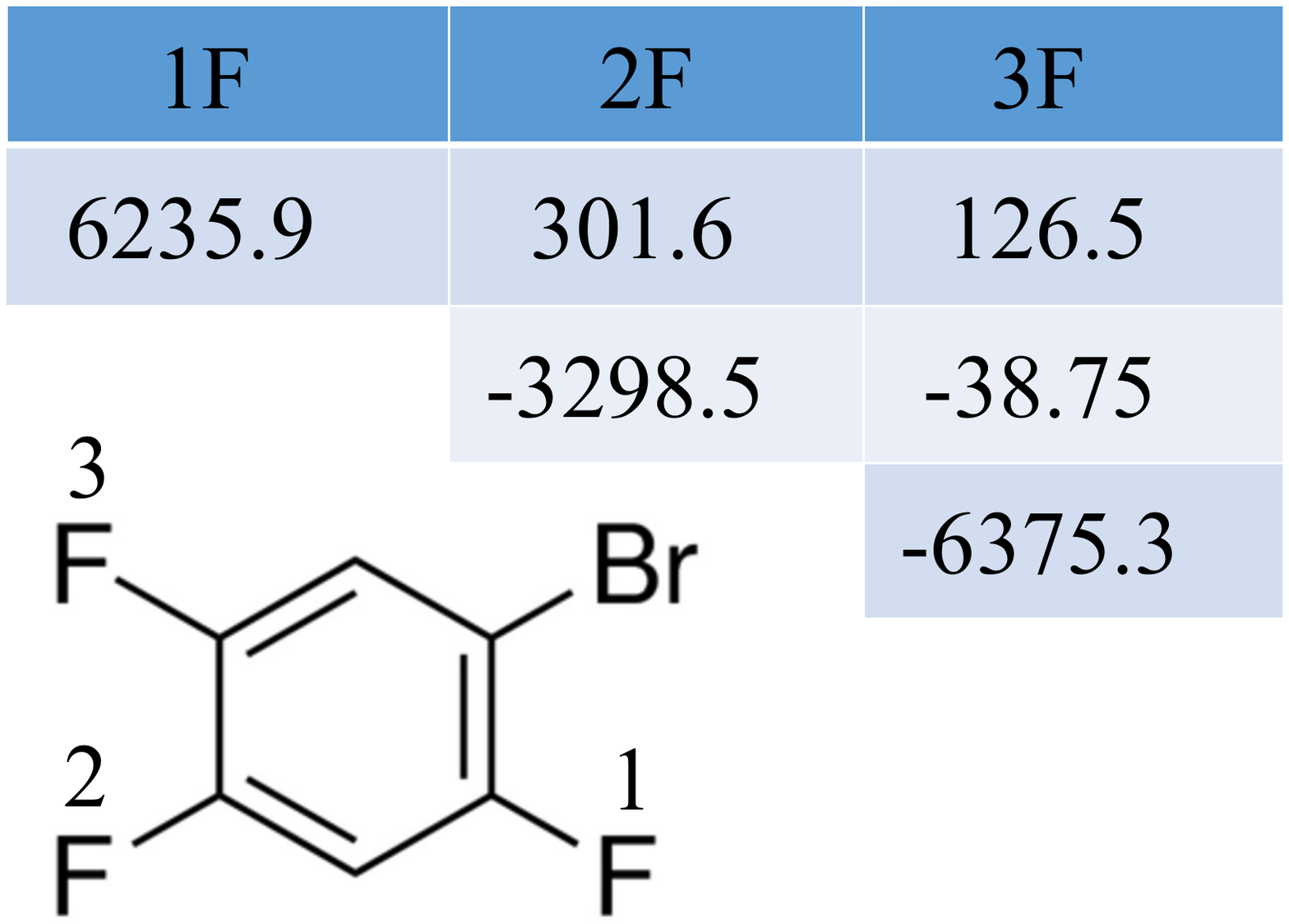}}%
    \qquad
    \subfloat[]{
  \includegraphics[trim=0cm 0cm 0cm 0.7cm,width=8cm,clip=]{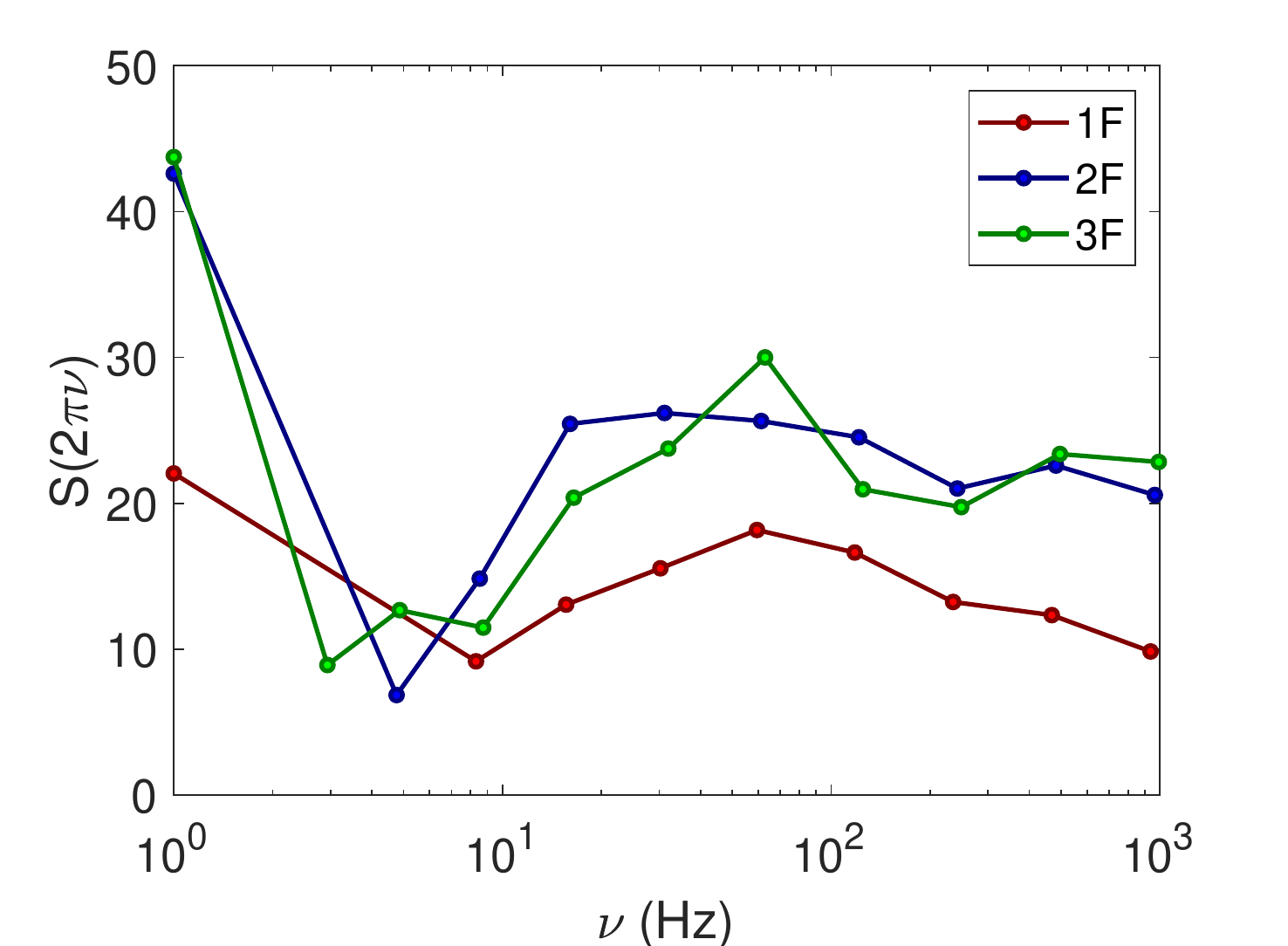}} %
    \caption{ (a) Molecular structure of BTFBz showing three $^{19}\mbox{F}$ nuclei and the table of resonance offsets $\Omega_k/(2\pi)$ (diagonal elements) and indirect spin-spin coupling constants $J_{kl}$ (off-diagonal elements) in Hz. (b) Experimentally obtained local noise spectrum for each $^{19}$F spin.}%
    \label{threespin}%
\end{figure}

Now we  demonstrate an experimental utilization of SAGRAPE by generating qubit-selective $\pi$ pulses and employ them to perform noise spectroscopy. For this purpose, we consider the three $^{19}$F spins of 1-bromo-2,4,5-tri
fluorobenzene (BTFBz) that is partially oriented in the liquid crystal N-(4-methoxybenzylidene)-4-butylaniline (MBBA). 
We applied spin-decoupling to remove the effects of the two hydrogen spins.
The  resonance offsets $\Omega_k/(2\pi)$ and  indirect spin-spin coupling constants $J_{kl}$ of the $^{19}$F spins  are shown in the inset table of Fig. \ref{threespin} (a).
 The noise spectroscopy  experiments  were  carried  out  in a  Bruker  500  MHz  NMR  spectrometer  at an ambient temperature of 300  K. 

Noise spectroscopy allows us to characterize the noise spectral density function $S(\nu)$ \cite{Noisespec,Noisespectros} as a function of the noise frequency $\nu_\delta$. The experiment involves measuring the  decay time-constant $T_2(\delta)$ of the magnetization with a set of CPMG sequences each with a specific inter-$\pi$ pulse delay $\delta$, that allows us to sample the noise spectrum
\begin{align}
S(2\pi\nu_\delta) \sim  \frac{\pi^2}{4 T_2(\delta)}
\end{align}
at frequencies $\nu_\delta = 1/(2\delta)$
\cite{Noisespec}.  
Understanding noise spectra is helpful to develop methods that protect quantum coherence against the environmental noise \cite{Biercuk}.

Carrying out the local noise-spectroscopy for each $^{19}$F spin needs a set of CPMG sequences involving spin-selective $\pi$ pulses.  Since a train of such $\pi$ pulses are used for finding the noise amplitude at each noise frequency, the cumulative pulse errors need to be small, thus necessitating the construction of high-fidelity $\pi$ pulses.

Using SAGRAPE-50 ($\kappa = 50)$, we generated three spin-selective $\pi$ pulses, one for each of the three $^{19}$F spins.
Each sequence was of total duration 360 $\upmu$s discretized into 360 equal time segments, and had an average fidelity over 0.99 for the RFI parameters $r_{m} = [0.8,1,1.2]$ and $p_m = [0.2,0.6,0.2]$ respectively.  
Fig. \ref{threespin} (b) displays the local noise spectrum for each of the three $^{19}$F spins.  These noise spectra are not only helpful in understanding the environment surrounding the spins, but also to generate  noise-resilient quantum controls tailored for them.
For the short-pulses discussed here, we may ignore the pulse errors occurring due to external noises.  However, for a long control sequence, errors gradually accumulate.  In the following section, we describe a general method to train the control algorithm against the external noises. 

\section{The robust simulated-annealing GRAPE (RSAGRAPE)}
\label{sec:Hamiltonian}
We now describe generating a control sequence that is robust against the dephasing noise, which generally is the predominant process limiting the coherence time of quantum systems.  In order to train the optimization algorithm against dephasing noise, we introduce an additional term in the Hamiltonian of Eq. \ref{eq:HandHrf}
\begin{align}
H^m(j) 
&= H_0 + H^m_\mathrm{RF}(j) + H^{m}_\mathrm{noise}(j)
\nonumber \\
\mbox{where,}~ H^m_\mathrm{noise}(j) &= 2\pi \eta^{m}(j) H_z.
 \end{align}

Here $\eta^{m}(j) \in [-\zeta/2,\zeta/2]$ is chosen from an uniform random distribution of range $\zeta$.  Optimizing in the presence of such a random phase-noise renders the control sequence robust against the dephasing effects of the environment.   We incorporate this technique of making robust controls with SAGRAPE algorithm explained in  section \ref{sec:Simuanneal} to create the RSAGRAPE (robust SAGRAPE) algorithm.
Additionally, we can introduce other decoherence-suppression methods such as dynamical decoupling. Here we integrate CPMG pulses within the control sequence as explained below.

\begin{figure}
	  \subfloat[]{
      \includegraphics[trim=0cm 0cm 0cm 0cm,width=8.5cm,clip=]{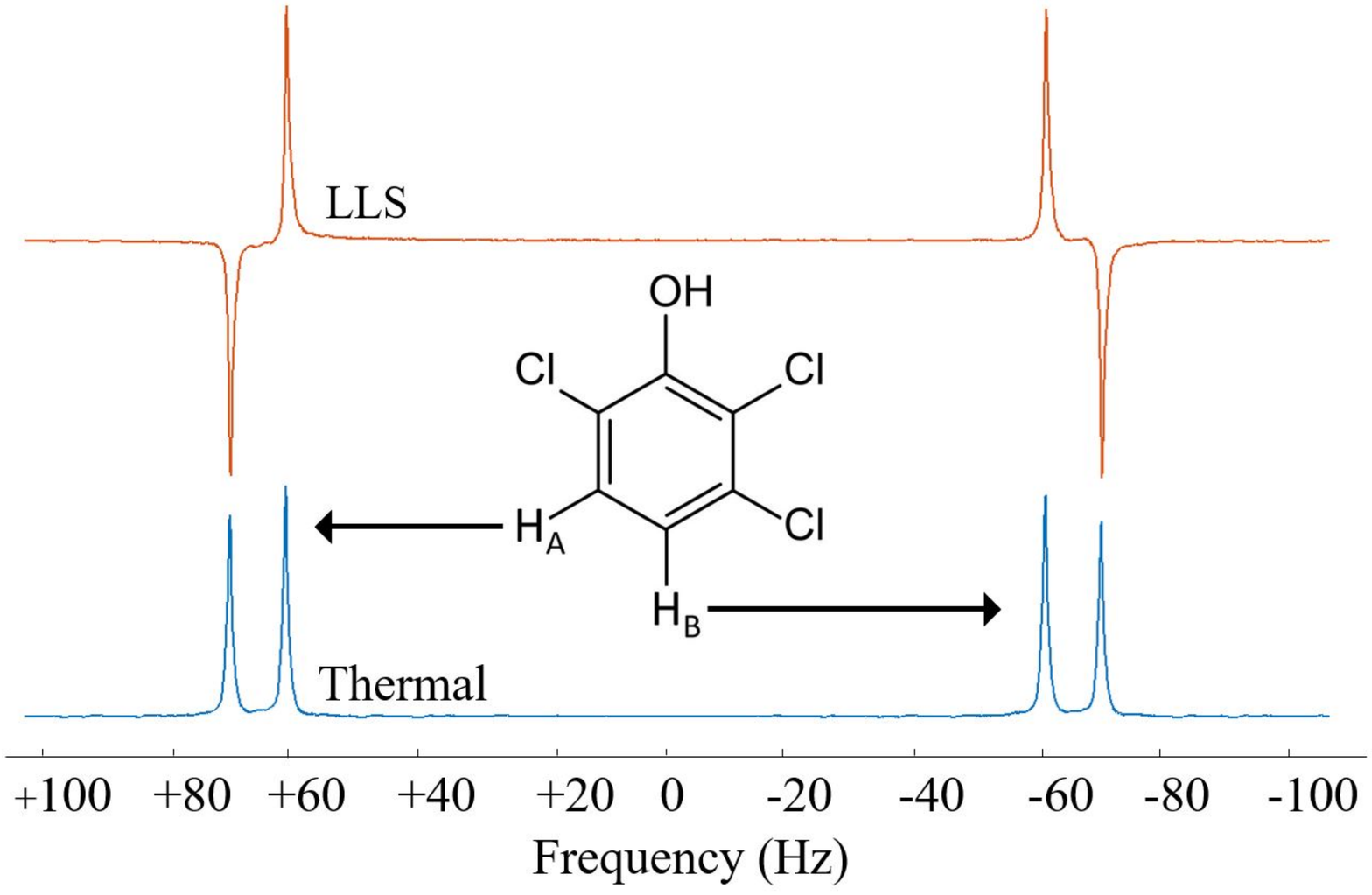}
      }
   	\qquad
   	 \subfloat[]{
   	% \vspace{+1em}
   	\includegraphics[trim=0cm 1.0cm 0cm 1cm,width=8.5cm,clip=]{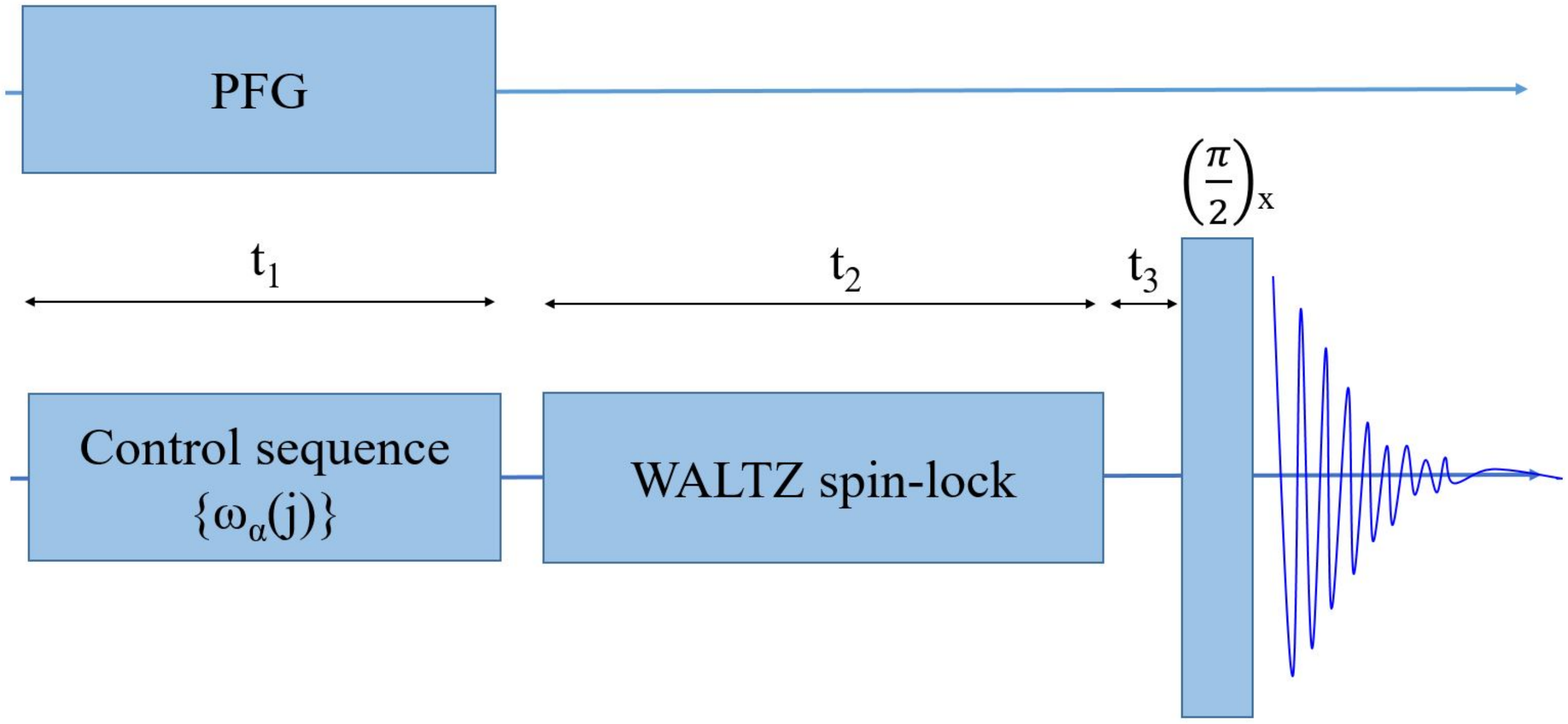}}
   \caption{(a) Molecular structure of TCP, its reference spectrum (blue), and its LLS spectrum (red). (b) The NMR pulse sequence for preparing, storing, and detecting the singlet order.  A pulsed-field-gradient (PFG) with amplitude randomly varying with time is used to induce phase noise.}
   	\label{fig:tcp}
   \end{figure}

We demonstrate the state controllability of RSAGRAPE  by preparing controls $\{\omega_{\alpha}(j)\}$  for two $^1$H nuclear spins  of 2,3,6 trichlorophenol (TCP) dissolved in dimethyl sulphoxide-D6 (DMSO). The molecular structure and $^1$H reference spectrum (in blue) are shown in Fig. \ref{fig:tcp} (a) with the two protons labelled as H$_A$ and H$_B$.
For this system, the difference in the resonance offsets $|\Omega_1-\Omega_2|/(2\pi)$ is 127.4 Hz and the indirect spin-spin coupling constant $J_{12}$ is 8.8 Hz.  The spin-lattice relaxation time constants (T$_1$) for the two spins obtained from the inversion recovery experiment were 5.5 s and 5.6 s for the spins H$_A$ and H$_B$ respectively \cite{Pushpull}.

Our objective is to generate a control sequence $\{\omega_\alpha(j)\}$ that transforms the state of the spin-system from thermal equilibrium to the long-lived singlet state (LLS) 

\begin{align}
I_{z1} + I_{z2}
\stackrel{\{\omega_\alpha(j)\}}{\longrightarrow}-\mathbf{I}_{1}\cdot \mathbf{I}_{2}.
\end{align}

While most non equilibrium quantum states decay to thermal states with a spin-lattice relaxation time constant $T_1$, the long-lived spin states can retain their spin-order for durations much longer than $T_1$ \cite{LevittSingletPRL2004}.
Buoyed by a number of interesting applications in several fields including spectroscopy, medical imaging, as well as quantum information \cite{roy2010initialization}, LLS has recently gained a significant attention \cite{pileio2020long}.  Typically, it takes a sequence longer than $1/(2J)$ to prepare LLS, during which time the noise can cause significant effects.   

\begin{figure}
	\subfloat[]{\includegraphics[trim=0cm 0.2cm 1cm 1cm,width=7.5cm,clip=]{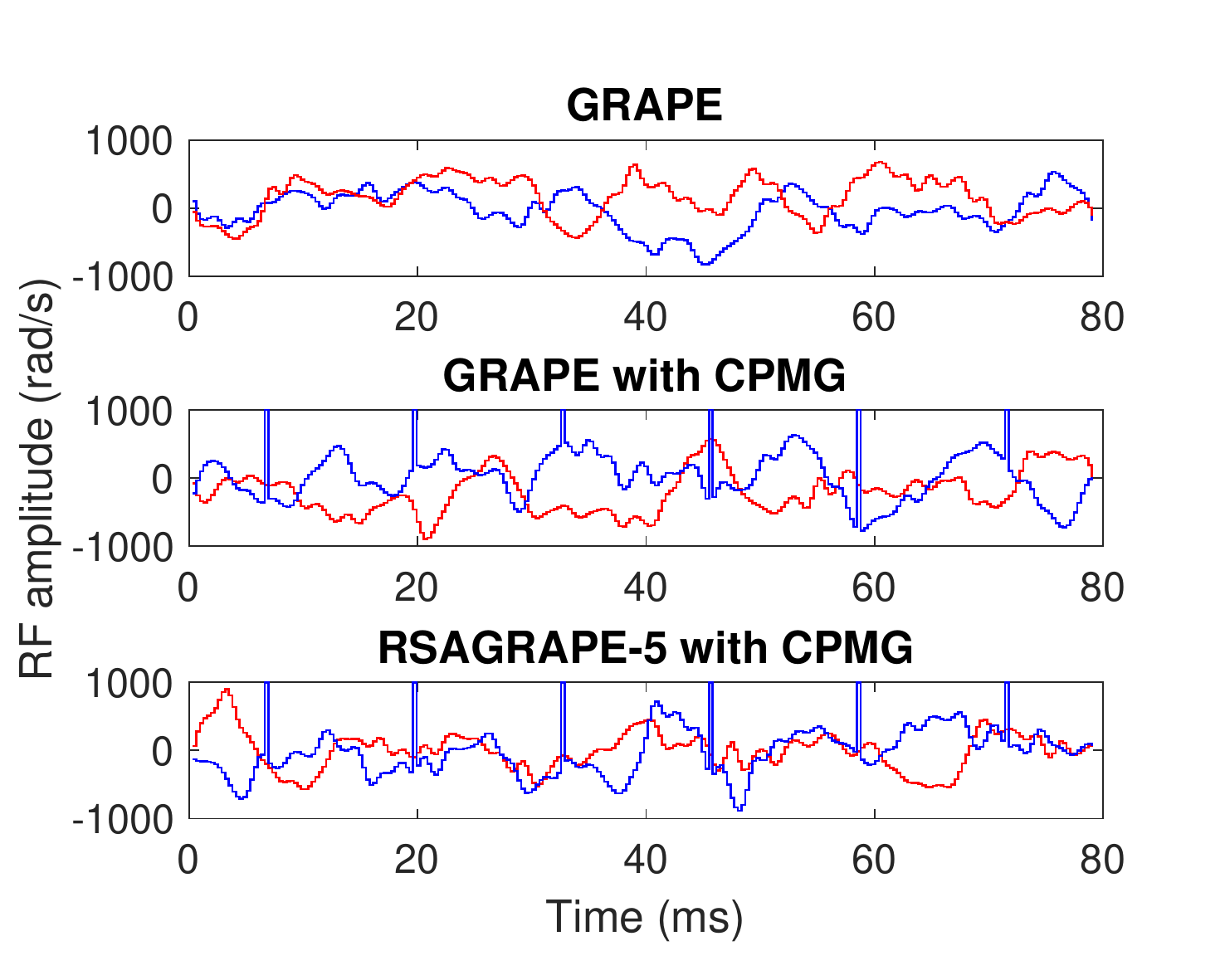}} 
   	\qquad
	 \subfloat[]{\includegraphics[trim=0cm 0cm 0cm 0.7cm,width=8cm,clip=]{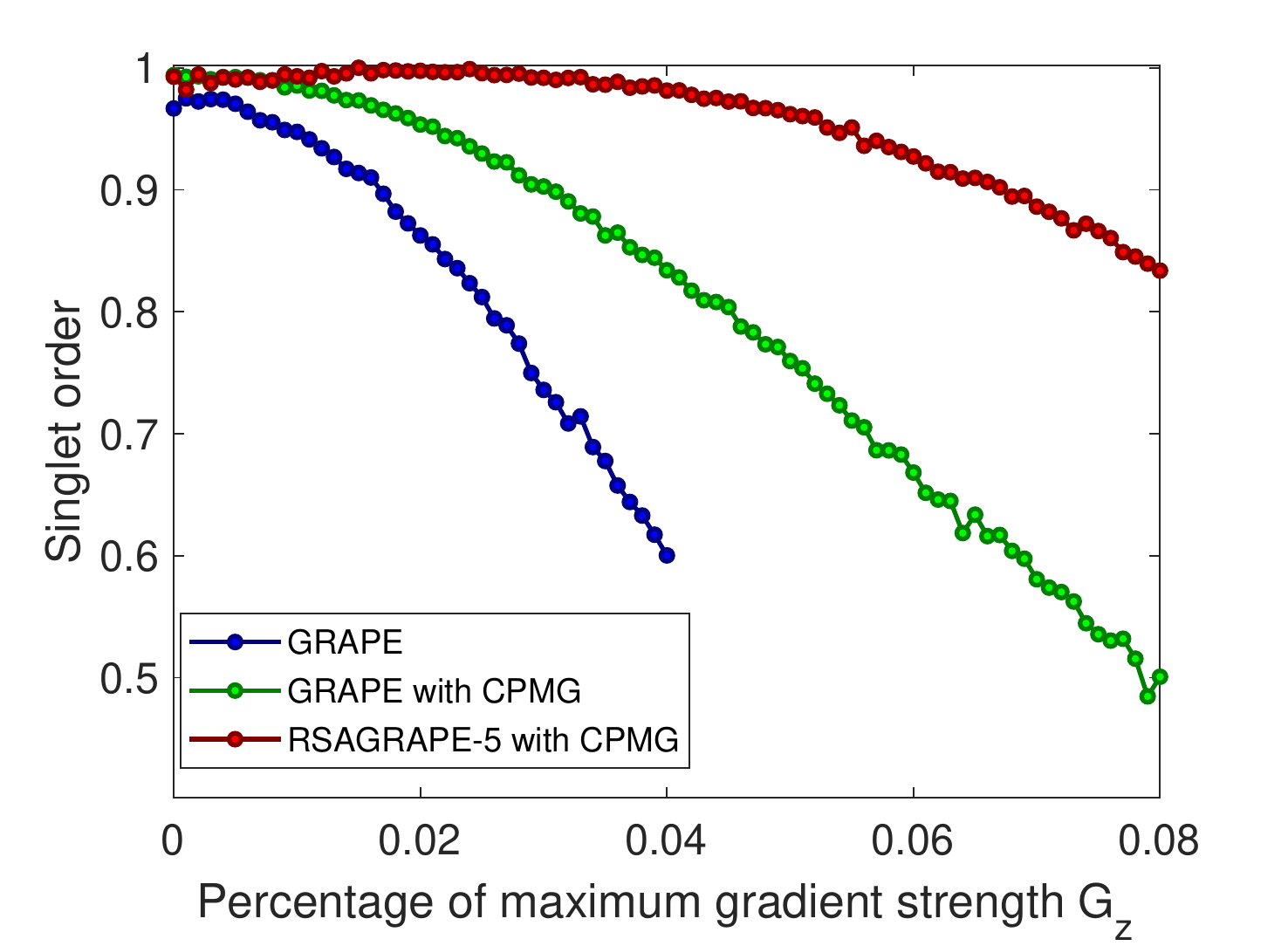}}
\caption{(a) The RF components $\omega_x(j)$ (blue, in rad/s) and $\omega_y(j)$ (red, in rad/s) corresponding to the three control sequences obtained with GRAPE, GRAPE with CPMG, and RSAGRAPE-5 with CPMG plotted versus time.  The latter two sequences are integrated with six CPMG $\pi$ pulses each of amplitude +9941 rad/s, but the latter two plots are cropped at $\pm$1000 rad/s for the visibility purpose. (b) The experimentally observed singlet order versus the maximum gradient strength $G_z$ (in $\%$) for the three control sequences shown in (a).
  	}  		\label{fig:threepulse}
   \end{figure}

We prepared the following three control sequences:
(i) GRAPE, (ii) GRAPE with CPMG, and (iii) RSAGRAPE-$\zeta$ with CPMG. Both (ii) and (iii) are integrated with six CPMG pulses.  The robust sequence RSAGRAPE-5 is generated using the RSAGRAPE algorithm with the noise parameter $\zeta = 5$ Hz and SA iterations $\kappa = 10$.
Each sequence is of duration $t_1=79$ ms, uniformly discretized into 250 segments and had an average fidelity over 0.99 with same RFI parameters $r_{m}=[0.9,1.0,1.1]$ and $p_m = [0.2,0.6,0.2]$.
The three sequences are plotted in Fig. \ref{fig:threepulse} (a).

The robustness of the three sequences are tested experimentally again in the Bruker 500 MHz spectrometer at an ambient temperature of 300 K.
The NMR pulse-sequence for preparing, storing, and measuring the singlet order is shown in Fig. \ref{fig:tcp} (b).  
After preparing LLS using each of the control sequences of duration $t_1$, we stored the LLS under WALTZ-16 spin-lock of 2 kHz amplitude for duration $t_2$.
Finally, we converted LLS into observable single-quantum magnetization with a free-evolution duration $t_3 = 1/(4J)$ followed by a $(\pi/2)_x$ pulse.  The resulting spectrum consists of a characteristic down-up-up-down spectrum as shown in the red trace of Fig. \ref{fig:tcp} (a).
By varying the storage duration $t_2$, we  estimated the life-time of the singlet-order to be 25 s, which is more than four times the $T_1$ values of the two $^1$H spins, thus confirming the preparation of LLS state.

We now experimentally compare the robustness of the three sequences against the dephasing noise.
To systematically control the dephasing noise, we introduced a pulsed-field gradient (PFG) $G_z$ that applies a spatial inhomogeneity along the direction of the Zeeman field, i.e., z-axis
(see Fig. \ref{fig:tcp} (b)). The amplitude of the PFG was randomly varied over time within a range $[-G_z,G_z]$ in each experiment.  In subsequent experiments, the amplitude $G_z$ was  systematically increased from 0 to 0.08\%, where 100\% refers to 50 G/cm.  In solution state, the nuclear spins are subjected to a controlled decoherence under the combined effects of random PFG as well as the random molecular motion due to translational diffusion.  Finally, we monitored the singlet-order by measuring the absolute area of the singlet spectrum obtained after a storage time $t_2 = 10$ s.  The experimental results are shown in Fig. \ref{fig:threepulse} (b).  Here all the data are normalized w.r.t. a common data corresponding to the highest singlet order.  It is clear that the simple GRAPE sequence (blue)  decays rapidly with the dephasing noise. 
The singlet order of the GRAPE sequence is lowest even at $G_z=0$, indicating that it is affected by the intrinsic noise of the spin system and the NMR setup.
On the other hand, GRAPE sequence integrated with CPMG pulses (green) performs relatively better.  In fact, it has the highest singlet-order in the absence of external dephasing, i.e., $G_z = 0$.  However, it is evident that RSAGRAPE-5 with CPMG (red) is the most robust sequence, that maintains high singlet-order for a wide range of dephasing noise.  Thus, it clearly establishes the superiority of the robust state control sequence generated by the RSAGRAPE algorithm in combating the dephasing noise. 

  %%%%%%%%%%%%%%%%%%%%%%%%%%%% Conclusion %%%%%%%%%%%%%%%%%%%%%%%%%%%%%%%%%%%%%%%%%%%%%%%%

\section{Conclusions}
\label{sec:conclusions}
Quantum control, which is crucial for realizing quantum technologies, is limited by two key factors (i) convergence efficiency of the optimization algorithms, and (ii) the robustness of the control sequence against external noises.  In this work, we address both of these factors.  

First, we combined simulated annealing (SA) with the commonly used gradient ascent algorithm (GRAPE) to realize a hybrid algorithm (SAGRAPE).  Our numerical analysis confirms that the convergence efficiency of the SAGRAPE algorithm is significantly improved over the GRAPE algorithm.  As a demonstration of an experimental application, we used the SAGRAPE algorithm to generate spin-selective $\pi$ pulses for a homonuclear three-spin NMR system and obtained their local noise spectra.  

Secondly, we proposed a general method to obtain noise-resilient control sequences by optimizing the control sequences in the presence of a noisy field.  In particular, we designed the RSAGRAPE (robust SAGRAPE) algorithm which generates robust control sequences against dephasing noise.  
Additionally, we incorporate CPMG pulses along with the control sequence which enhances their robustness against external noise.
By  experimentally comparing the preparation efficiency of long-lived singlet states in the presence of a controlled external noise, we confirm the superiority of the RSAGRAPE sequence over the GRAPE sequences.

In principle, the convergence efficiency of the hybrid algorithm can be further improved by incorporating more advanced variants of simulated annealing such as adaptive simulated annealing (ASA) \cite{adaptivesimuanneal}. We believe that such hybrid algorithms will play an important role in the future as we attempt to control larger quantum systems with higher precision.

\section*{Acknowledgements}
The authors gratefully acknowledge NMR hardware assistance from Nitin Dalvi. PB acknowledges support from Prime Minister Research Fellowship. TSM acknowledges funding from DST/ICPS/QuST/2019/Q67. 

\bibliography{bibliography}{}
\bibliographystyle{apsrev4-1}

\end{document}